\documentclass[fleqn,usenatbib]{mnras}

\usepackage{newtxtext,newtxmath}
\usepackage[T1]{fontenc}
\usepackage{balance}
\usepackage{flushend}

\DeclareRobustCommand{\VAN}[3]{#2}
\let\VANthebibliography\thebibliography
\def\thebibliography{\DeclareRobustCommand{\VAN}[3]{##3}\VANthebibliography}

\usepackage{graphicx}	% Including figure files
\usepackage{amsmath}	% Advanced maths commands
\usepackage{cuted}
\usepackage{caption}

\newcommand{\warwick}{1}
\newcommand{\hamburg}{2}
\newcommand{\sheffield}{3}
\newcommand{\iac}{4}
\newcommand{\sheffieldrse}{5}
\newcommand{\oklahoma}{6}
\newcommand{\colorado}{7}
\newcommand{\cork}{8}
\newcommand{\potsdam}{9}

\title[Spin-down of the White Dwarf in J1912-4410]{Determining the Spin-down of the White Dwarf in J191213.72-441045.1 using Optical Time-Series Photometry}

\author[C. J. Hodge et al.]{
Callum J. Hodge,$^{\warwick}$\thanks{E-mail: Callum.Hodge@warwick.ac.uk},
Ingrid Pelisoli$^{\warwick}$\thanks{E-mail: Ingrid.Pelisoli@warwick.ac.uk},
A.~J. Brown$^{\hamburg}$,
N. Castro Segura$^{\warwick}$,
V.~S. Dhillon$^{\sheffield,\iac}$,
M.~J. Dyer$^{\sheffield,\sheffieldrse}$,\newauthor
J.~A. Garbutt$^{\sheffield}$,
M.~J. Green$^{\oklahoma,\colorado}$
D. Jarvis$^{\sheffield}$,
M. R. Kennedy$^{\cork}$,
P. Kerry$^{\sheffield}$,
S.~P. Littlefair$^{\sheffield}$,
J. McCormac$^{\warwick}$,\newauthor
J. Munday$^{\potsdam}$,
S.~G. Parsons$^{\sheffield}$,
E. Pike$^{\sheffield}$,
D.~I. Sahman$^{\sheffield}$,
A.~Yates$^{\sheffield}$,
T.~Zivave$^{\warwick}$
\\
$^{\warwick}$Department of Physics, University of Warwick, Gibbet Hill Road, Coventry, CV4 7AL, UK\\
$^{\hamburg}$Hamburger Sternwarte, University of Hamburg, Gojenbergsweg 112, 21029 Hamburg, Germany\\
$^{\sheffield}$Astrophysics Research Cluster, School of Mathematical and Physical Sciences, University of Sheffield, Sheffield, S3 7RH, United Kingdom\\
$^{\iac}$Instituto de Astrof\'{i}sica de Canarias, E-38205 La Laguna, Tenerife, Spain\\
$^{\sheffieldrse}$Research Software Engineering, University of Sheffield, Sheffield, S1 4DP, United Kingdom\\
$^{\oklahoma}$Homer L. Dodge Department of Physics and Astronomy, University of Oklahoma, 440 W. Brooks Street, Norman, OK 73019, USA\\
$^{\colorado}$JILA, University of Colorado and National Institute of Standards and Technology, 440 UCB, Boulder, CO 80309-0440, USA\\
$^{\cork}$School of Physics, University College Cork, Cork, T12 K8AF, Ireland\\
$^{\potsdam}$Institute for Physics and Astronomy, University of Potsdam, Karl-Liebknecht-Str. 24/25, 14476 Potsdam, Germany\\
}

\date{Currently under peer-review. Received in original form 2026 September 14.}

\pubyear{\the\year{}}

\begin{document}
\label{firstpage}
\pagerange{\pageref{firstpage}--\pageref{lastpage}}
\maketitle

\begin{abstract}
J191213.72-441045.1 (J1912-4410) is the second white dwarf pulsar to be discovered. These binary systems consist of a white dwarf primary with a cool companion and undergo pulsed emission from radio to X-ray wavelengths. The prototype of the class, AR Scorpii (AR Sco), exhibits a lack of accretion, which led to suggestions that the spin-down of the white dwarf is powering the system's non-thermal emission. However, J1912-4410 displays possible signatures of accretion, which allows for the possibility that its emission may operate via a different mechanism to AR Sco. To investigate this, we determined the first estimate for the spin period derivative of J1912-4410 from optical photometric data. This was carried out using ULTRACAM observations in multiple bandpasses spanning the last four years. Our estimate for the spin period derivative of $4.13(8)\times10^{-12}$\,s\,s$^{-1}$ is consistent across the $u_s$, $g_s$ and $r_s$ bandpasses of ULTRACAM. The corresponding spin-down power of $9^{+6}_{-5}\times10^{25}$\,W is approximately equal to previous estimates for the system's bolometric luminosity. This suggests that the non-thermal emission of J1912-4410 is not solely powered by its spin-down, as near-perfect efficiency would be required. Additionally, our spin period derivative indicates that J1912-4410 is unlikely to be consistent with the synchronar model proposed to explain the emission of AR Sco, given previous upper limits for the white dwarf's magnetic field. Our findings imply that J1912-4410 has a distinct emission mechanism compared to AR Sco, which, if true, means that white dwarf pulsars are a heterogeneous class of binary systems. 
\end{abstract}


%%%%%%%%%%%%%%%%%%%%%%%%%%%%%%%%%%%%%%%%%%%%%%%%%%

%%%%%%%%%%%%%%%%% BODY OF PAPER %%%%%%%%%%%%%%%%%%

\section{Introduction}
\label{sec:Introduction}

White dwarf (WD) pulsars are a rare class of binary systems consisting of a magnetic WD primary and a cool companion. These systems are characterised by their periodic, pulsed emission (modulated on the beat or spin period of the WD binary system) which spans radio \citep[]{2018A&A...611A..66S} to X-ray \citep[]{2018ApJ...853..106T} wavelengths. This, along with a lack of mass accretion from the donor star \citep[]{2025MNRAS.543.2116C}, makes them distinct from other known types of WD binary systems, such as cataclysmic variables (CVs). Despite this distinction, there are suggestions that WD pulsars may have evolutionary links to CVs and other WD binary systems. For example, WD pulsars have been proposed as a short-lived, intermediary phase in the evolution of accreting WD binaries, where the emergence of a magnetic field causes the system to become temporarily detached \citep[]{2021NatAs...5..648S}. The importance of WD pulsars also extends much further than WD binary evolution. For example, their characteristic lack of accretion and strong magnetic fields \citep[]{2016Natur.537..374M} make WD pulsars an ideal laboratory for studying the generation and decay of magnetic fields in binaries and how this influences angular momentum transfer between the two components. As a result, our lack of knowledge of WD pulsars represents an opportunity to advance our understanding of binary systems, astrophysical magnetic fields, and radiation mechanisms.

The prototype of the WD pulsar class, AR Scorpii (henceforth AR Sco), is a binary consisting of a rapidly-rotating WD (spin period of $1.95$\,min) and a companion with spectral type M4.5, orbiting at a period of $3.56$\,h \citep[]{2016Natur.537..374M}. The high fraction of optical linear polarisation ($\approx40$\,\%) suggests that its pulses are the result of synchrotron emission \citep[]{2017NatAs...1E..29B}. The radio emission, on the other hand, is strongly circularly polarised (up to 30\%), hinting at the existence of a cyclotron emission component \citep[]{2018A&A...611A..66S}. Weak X-ray emission ($\approx5\times10^{30}$\,erg\,s$^{-1}$), as well as a lack of flickering or broadened emission lines, disfavour an accretion-powered emission scenario \citep[]{2016Natur.537..374M, 2017NatAs...1E..29B}. Instead, it is believed that the spin-down of the WD powers the non-thermal emission released as magnetic dipole radiation (primarily in radio wavelengths), analogous to a traditional pulsar \citep[]{2016Natur.537..374M, 2017NatAs...1E..29B}. \cite{2022MNRAS.516.5052P} used high-speed optical photometric data spanning almost 17 years to find a frequency derivative for the WD in AR Sco of $-4.74(8)\times10^{-17}$\,Hz\,s$^{-1}$, an improvement on estimates from \cite{2018AJ....156..150S} and \cite{2020MNRAS.496.4849G}, which is sufficient to power the observed non-thermal emission.

\begin{table}
	\centering
	\caption{Basic meta data of ULTRACAM $u$, $g$, $r$ and $i$ band observations. We omit the exposure times from this table, as these are approximately equal to the cadences for each filter due to the minimal dead time ($24$\,ms) of ULTRACAM.}
	\label{tab:Observations}
	\begin{tabular}{ccccc} % four columns, alignment for each
		\hline
		UTC Start Date & Length (h) & Filters & \multicolumn{2}{c}{Cadence (s)}\\
		& & & $u$ & $g/r/i$\\
		\hline
		27 Apr 2022 & 0.5 & $ugi$ & 18.6 & 6.2\\
		06 Jun 2022 & 4.6 & $ugi$ & 32.0 & 8.0\\
		07 Jun 2022 & 4.0 & $ugi$ & 9.1 & 3.0\\
        17 Sep 2022 & 0.9 & $ugr$ & 12.1 & 4.0\\
        23 Sep 2022 & 3.9 & $ugr$ & 8.5 & 2.8\\
        29 Mar 2023 & 2.1 & $ugi$ & 9.0 & 3.0\\
        23 Apr 2023 & 1.4 & $ugr$ & 10.1 & 3.4\\
        27 Apr 2023 & 0.5 & $ugi$ & 8.5 & 2.8\\
        03 Sep 2023 & 4.6 & $ugr$ & 32.0 & 8.0\\
        18 Oct 2023 & 0.8 & $ugi$ & 10.1 & 3.4\\
        01 May 2024 & 2.8 & $ugi$ & 10.1 & 3.4\\
        01 May 2024 & 1.6 & $ugi$ & 10.1 & 3.4\\
        03 May 2024 & 2.0 & $ugr$ & 9.1 & 3.0\\
        04 May 2024 & 2.5 & $ugr$ & 16.1 & 4.0\\
        07 May 2024 & 0.9 & $ugi$ & 10.1 & 3.4\\
        28 Mar 2025 & 2.6 & $ugr$ & 13.7 & 3.4\\
        30 Mar 2025 & 1.0 & $ugr$ & 10.1 & 3.4\\
        02 Apr 2025 & 1.8 & $ugr$ & 10.1 & 3.4\\
        08 Apr 2025 & 0.9 & $ugi$ & 10.1 & 3.4\\
        26 Apr 2025 & 1.3 & $ugi$ & 12.1 & 4.0\\
        27 Jun 2025 & 1.9 & $ugi$ & 10.1 & 3.4\\
        28 Jun 2025 & 2.3 & $ugr$ & 10.1 & 3.4\\
        26 Sep 2025 & 1.7 & $ugr$ & 13.5 & 3.4\\
        26 Sep 2025 & 1.3 & $ugr$ & 13.5 & 3.4\\
        01 Nov 2025 & 1.1 & $ugi$ & 11.0 & 3.7\\
        18 Mar 2026 & 0.6 & $ugi$ & 10.1 & 3.4\\
        19 Mar 2026 & 1.4 & $ugi$ & 9.2 & 3.1\\
        14 Apr 2026 & 1.3 & $ugr$ & 12.5 & 4.2\\
		\hline
        Average & 1.9 & -- & 12.6 & 3.8\\
        \hline
	\end{tabular}
\end{table}

However, the similarities to neutron star pulsars appear to end here. Unlike traditional pulsars, WDs do not possess the sufficient pair production conditions in their magnetospheres to generate the relativistic electrons and positrons that can produce beamed non-thermal emission \citep[]{2024ApJ...961..214R}. Thus, AR Sco must generate its non-thermal emission using a different mechanism. One proposed solution to this is the \textit{synchronar model} by \cite{2017ApJ...835..150K}, where the spin-down power is dissipated in the M-dwarf atmosphere via magnetic reconnection, and this accelerates the particles which produce non-thermal emission. This would mean that such emission originates from the M-dwarf atmosphere, without the need for continual mass transfer between the two binary components, and is modulated by the periodic sweeping of the WD magnetosphere across the surface of the M-dwarf. Additionally, a strong WD magnetic field is necessary to provide a sufficient torque on the companion M-dwarf. Despite the plausibility of the synchronar model, its success has been limited by an inability to gain accurate estimates for the magnetic field of AR Sco; measurements for the upper bound of the WD magnetic field of $\lesssim100$\,MG from \cite{2021ApJ...908..195G} using far-UV spectral analysis agree with estimates from the synchronar model, yet submillimetre observations from \cite{2025ApJ...986...78B} suggest the magnetic field of the WD could be as low as $\approx15$\,MG, hence challenging the model. Furthermore, these submillimetre observations suggest that the non-thermal emission of AR Sco originates from near the WD and not from the M-dwarf atmosphere, a direct contradiction of the synchronar model. Evidently, observations of AR Sco across multiple wavebands pose mixed results for verifying the synchronar model, meaning observations of other WD pulsars may be essential to learning what the exact emission mechanism of WD pulsars is, and whether it is common across multiple systems.

\begin{figure}
	\includegraphics[width=\columnwidth]{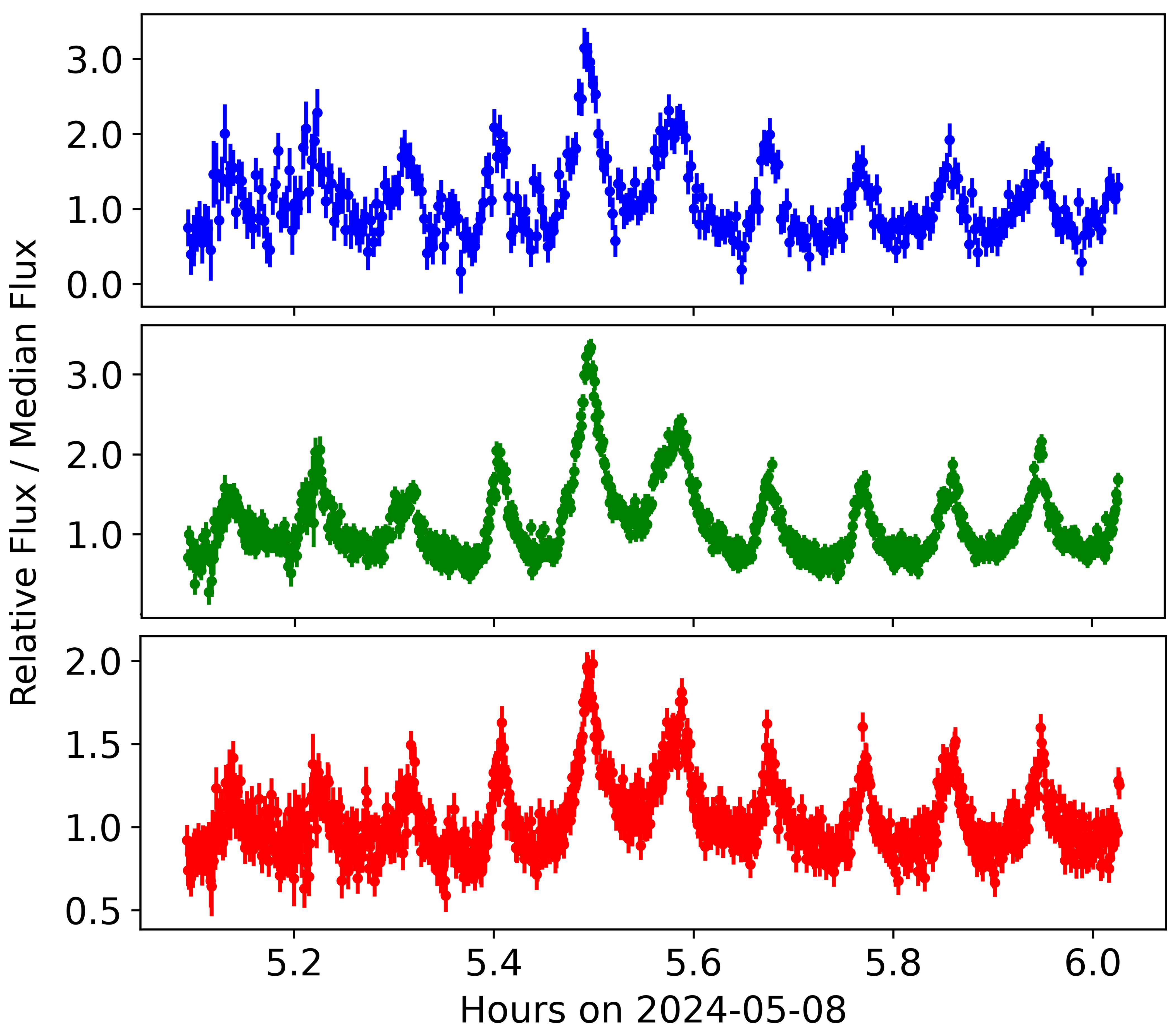}
    \caption{Lightcurve of J1912-4410 in the optical $u$ (top), $g$ (middle) and $i$ (bottom) bandpasses where the WD pulsar shows clear pulsed and periodic emission which is modulated by the spin frequency. Here, we plot the relative flux divided by the median flux of the observation night.}
    \label{fig:ExampleLightcurve}
\end{figure}

The second WD pulsar to be discovered was J191213.72-441045.1 \citep[henceforth J1912-4410,][]{2023NatAs...7..931P, Schwope2023}. Like AR Sco, J1912-4410 consists of a rapidly-spinning WD with an M-dwarf companion, a property, so far, shared by all WD pulsars \citep[]{2023NatAs...7..931P, 2025MNRAS.543.2116C}. The spin and orbital periods of J1912-4410 are longer than those of AR Sco, $5.30$\,min and $4.03$\,h respectively \citep[]{2023NatAs...7..931P}. Also similar to AR Sco, the system produces periodic emission over a broad range of wavelengths \citep[]{2023NatAs...7..931P}. However, the WD's estimated magnetic field of $\approx50$\,MG \citep[]{2024MNRAS.527.3826P} may, like the submillimetre observations from \cite{2025ApJ...986...78B}, be too weak to be explained by the synchronar model, depending on the WD's spin-down rate. In addition, X-ray observations of J1912-4410 from \cite{2023NatAs...7..931P} and \citet{Schwope2023} indicate possible signatures of residual accretion, meaning that, unlike AR Sco, J1912-4410 may be undergoing mass transfer from the M-dwarf companion onto the WD. Accretion could provide an additional source of energy to power the non-thermal emission that does not originate from the spin-down power. Both of these distinctions open up the possibility that WD pulsars are a heterogeneous class of WD binary systems with multiple, distinct mechanisms and energy sources for their characteristic, non-thermal emission. WD pulsars are already members of a scarce, possibly heterogeneous class of objects known as Long Period Radio Transients (LPTs): radio sources that exhibit uniform, periodic emission on timescales of seconds to minutes \citep[]{2026JHEAp..5200566R}. Since other members of this group have also been identified as possible WD pulsars, such as GPM J1839-10 \citep[]{2026NatAs..10..522H}, this further hints at the possibility that WD pulsars are a heterogeneous class. One piece of information that can give us greater insight into these matters is the spin-down of the WD in J1912-4410.

Here, we present the first estimate for the spin-down of the WD in J1912-4410 from high-speed optical photometric data. In finding this estimate, we compare the spin-down rate to that of AR Sco and verify whether this is consistent with the synchronar model, given the upper limit on the magnetic field of J1912-4410 established by \cite{2024MNRAS.527.3826P} of $\approx50$\,MG. Finally, we establish the spin-down power of J1912-4410 and determine whether this is sufficient to power the non-thermal emission.

\begin{figure}
	\includegraphics[width=\columnwidth]{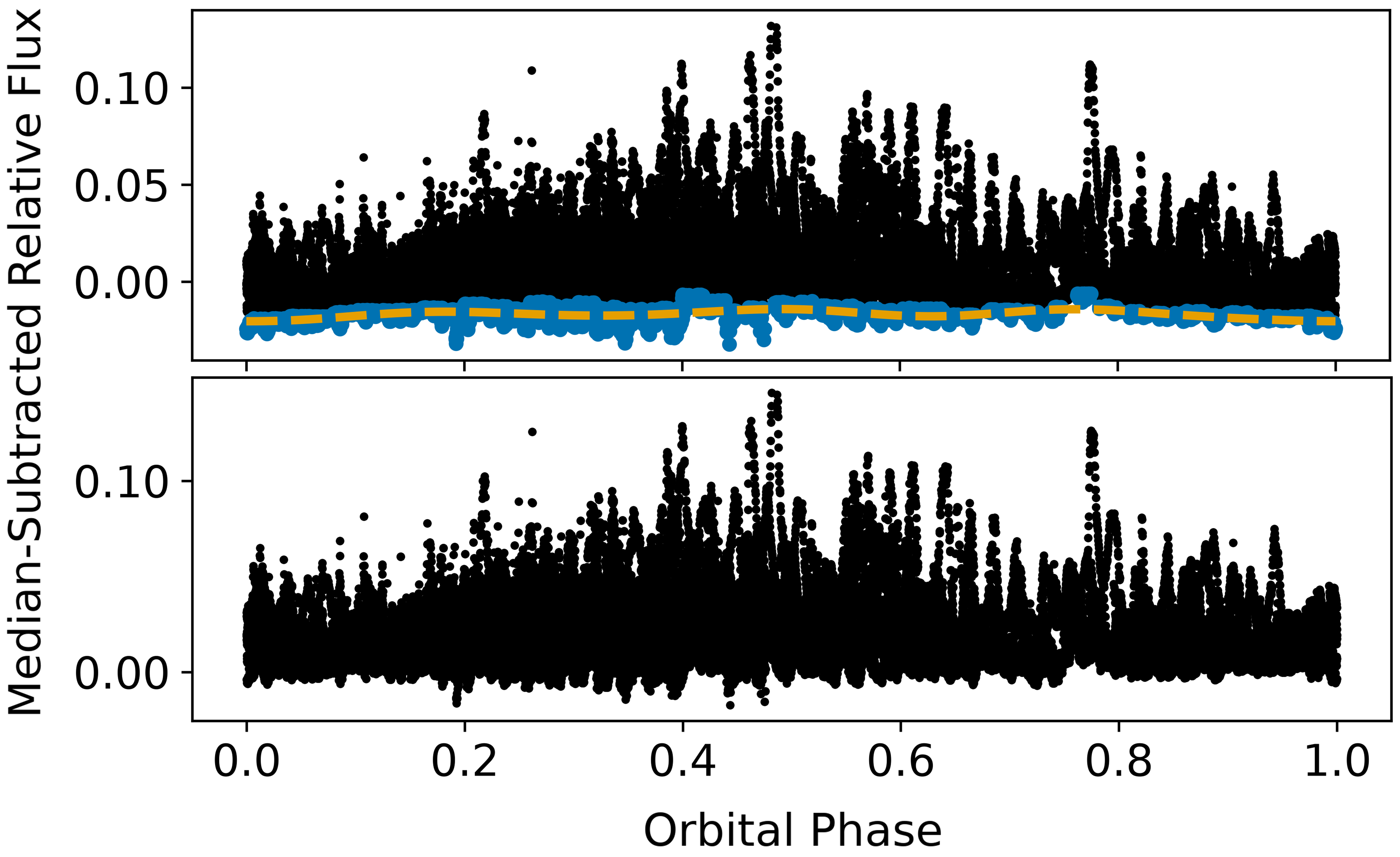}
    \caption{The $g$ filter lightcurve of J1912-4410 phase-folded onto the orbital period and median-subtracted using the median flux from each observing night. We divided the data into 50 equally-spaced orbital phase bins. The top panel shows the data with the lowest $7.5$\,\% of median-subtracted relative flux in each bin highlighted in blue, as well as a 10-term Fourier series fitted to this subset of the data displayed as a dashed orange line. The bottom panel then shows the orbital phase-folded lightcurve with the Fourier series subtracted from each data point.}
    \label{fig:OrbitalModulation}
\end{figure}

\section{Observations}
\label{Section: Observations}

We observed J1912-4410 using the ULTRACAM instrument \citep[]{2007MNRAS.378..825D}, mounted on the $3.5$\,m ESO New Technology Telescope (NTT) in Chile. Such observations were carried out in the $u_s$, $g_s$, $r_s$ and $i_s$ bandpasses, centred on $3526$\,\AA, $4732$\,\AA, $6261$\,\AA\ and $7672$\,\AA\ respectively. These filters are similar to those of the Sloan Digital Sky survey \citep[]{1996AJ....111.1748F}, but with a higher throughput. For simplicity, we drop the {\it s} subscript henceforth. Our data consisted of a total of 28 nights of observations (see Table~\ref{tab:Observations}), and thus we were able to observe how the timing of the pulsations of J1912-4410 changed over a time span of almost four years (2022--2026). It should also be noted that ULTRACAM observes simultaneously in three bands, but it cannot simultaneously observe in the $r$ and $i$ bands. Therefore, despite the observations in these bands still covering a comparable timespan to the $u$ and $g$ filters, we only observed J1912-4410 in the $r$ and $i$ bands for 13 and 15 nights respectively. Additionally, the errors in pulse arrival times for the $u$ filter are the greatest due to their high relative exposure times. For these reasons, the majority of our discussion centres around analysis of the $g$ filter data, the data with both low errors and a high number of data points. We used the HiPERCAM pipeline to reduce all data sets \footnote{https://github.com/HiPERCAM/hipercam}, applying bias and flat-field corrections and performing aperture photometry using the same constant comparison star as in \cite{2023NatAs...7..931P} (Gaia EDR3 6712712349514963072, $G=13.8$). Henceforth, we refer to the flux from J1912-4410 divided by the comparison star flux as the \textit{relative flux}. All recorded observation times underwent a Barycentric correction and were converted to a Modified Julian Date (MJD). An example lightcurve is diplayed in Fig.~\ref{fig:ExampleLightcurve}.

\begin{figure}
	\includegraphics[width=\columnwidth]{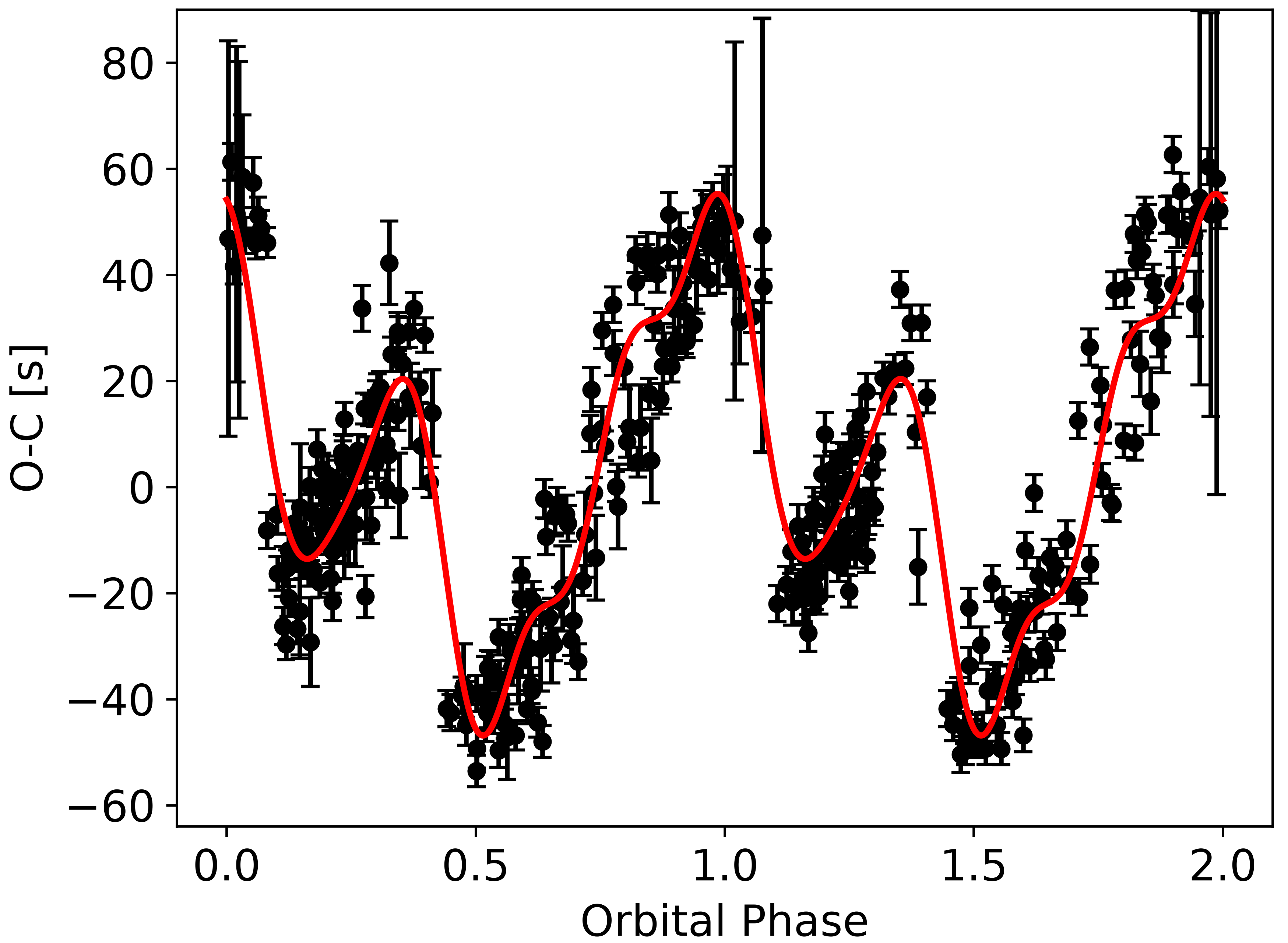}
    \caption{Observed minus calculated (O--C) residuals of identified arrival times of pulses from J1912-4410 in comparison to a quadratic ephemeris, phase-folded onto the orbital period. These residuals are displayed using $g$ filter data, yet the behaviour is similar for all filters used. The red line displays a 10-term Fourier series, which we used to model this behaviour and incorporate it into all quadratic ephemerides.}
    \label{fig:FourierSeries}
\end{figure}

\section{Data Analysis}
\label{Section: Data Analysis}
Before our analysis, we removed any major outliers from the data, which were identified using the median absolute deviation \citep[]{MAD}. From repeated trial and inspection of the data to check that valid data points were not being erroneously removed, we denoted outliers as having a median absolute deviation $>7$. Data points flagged by the HiPERCAM reduction pipeline as having bad counts either for J1912-4410 or the comparison star were removed. We also removed the first frame for each filter in each observing run, as this had an unreliable MJD (also flagged in the HiPERCAM reduction pipeline). 

We also sought to remove any variations in the lightcurve that were modulated on the orbital period, using the following method \citep[similar to][]{2018AJ....156..150S}. Data for each night were median-subtracted, then the entire data were orbital phase-folded \citep[using the orbital period from][]{2023NatAs...7..931P} and divided into $50$ equally-spaced bins. We then fitted a 10-term Fourier series to the lowest $7.5$\,\% of data in all bins (the top plot of Fig.~\ref{fig:OrbitalModulation}), and then subtracted this Fourier fit from all data points (the bottom plot of Fig.~\ref{fig:OrbitalModulation}). This is because the data points with the lowest $7.5$\,\% of median-subtracted relative flux in each bin were taken to be representative of the pure underlying orbital modulation, with no contribution to the pulses. Unexpectedly, we also found that the shape of this orbital modulation changed across the four years of our observations. We discuss this further in Appendix~\ref{appsec: orbital modulation}.

\begin{figure}
	\includegraphics[width=\columnwidth]{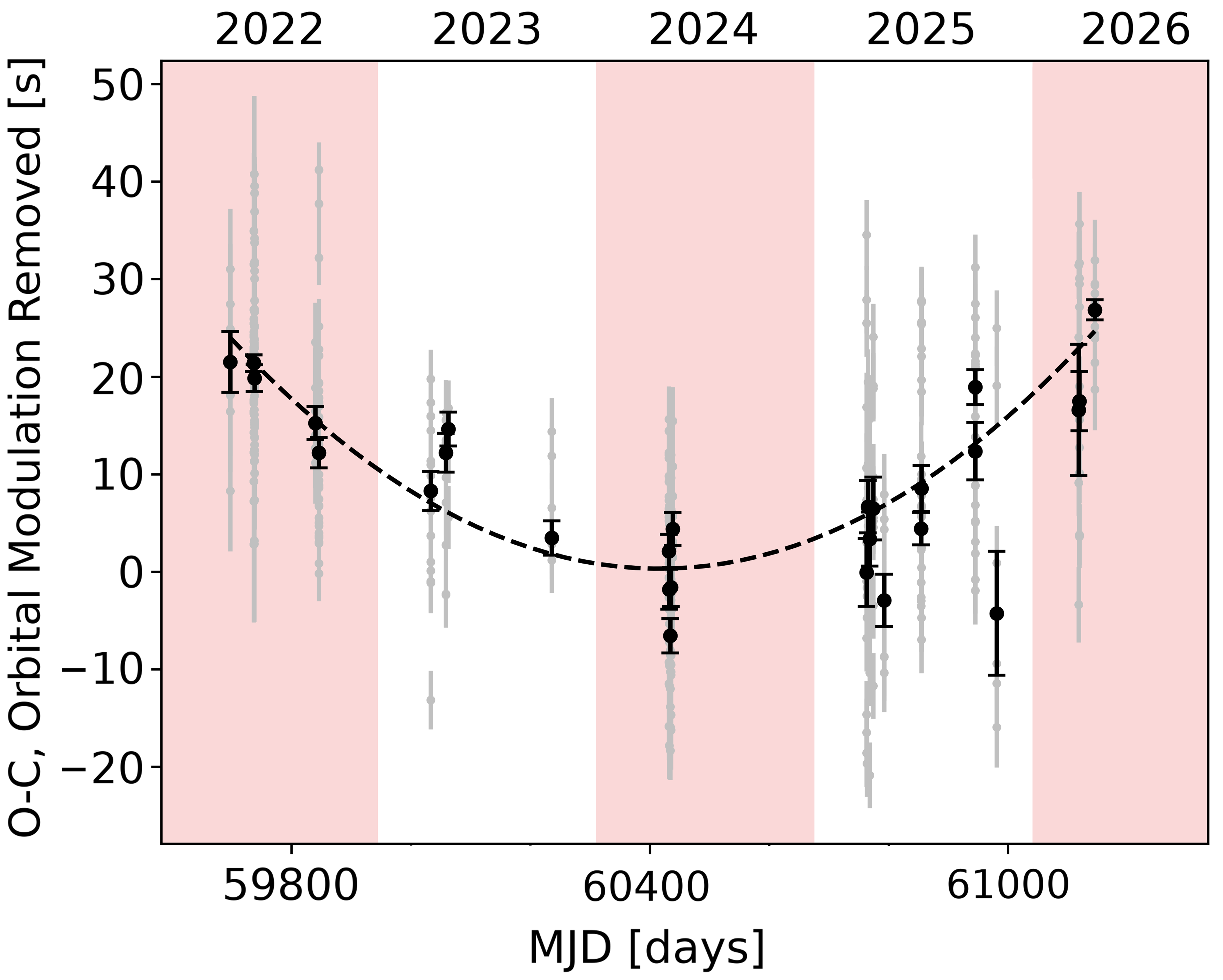}
    \caption{Residuals plot (observed minus calculated, O--C) of pulse arrival times after the linear ephemeris is subtracted and the residuals in Fig.~\ref{fig:FourierSeries} are corrected for using $g$ filter data. The grey data points display the raw data. We also bin the data according to the night of observation; the black data points display the mean O--C of each bin, with the error calculated from the standard error within each bin. The quadratic fit is plotted as a dashed line. Different years of observation are highlighted by alternating red and white background columns, with the years marked on top.}
    \label{fig:OminusCPlot}
\end{figure}

We were then in a position to determine the arrival times of each pulse from J1912-4410 from the reduced lightcurves. To do this, we employed the method of \cite{2022MNRAS.516.5052P}, used to determine the arrival times of pulses from AR Sco. We first determined the expected arrival times of each pulse using the linear ephemeris from \cite{2023NatAs...7..931P}, as well as a quadratic term, calculated using the order of magnitude of the beat period derivative of AR Sco from \cite{2022MNRAS.516.5052P}. It should be noted at this point that the dominant frequency for AR Sco at the time of discovery was the beat frequency \citep[]{2016Natur.537..374M}, unlike for J1912-4410 where this is the spin frequency \citep[]{2023NatAs...7..931P}, which is why we use AR Sco's beat period derivative to form our initial ephemeris. Our expected arrival times allowed us to form a window around each expected pulse, which was chosen to be $120$\,s ($60$\,s either side of the expected arrival time, chosen to maximise the number of clear peaks identified, from trial and inspection of the data). Then, for each window of data, we convolved this with a Gaussian, took the derivative of this cross-correlation function and identified where the derivative went from positive to negative, hence identifying the position of a candidate peak in the data. If this crossing was in between data points, we used linear interpolation to estimate the peak location. The standard deviation ($\sigma$) of the Gaussian function was chosen to minimise the root-mean-squared residuals of the quadratic ephemeris. After experimenting with values from $5$\,s to $40$\,s, the optimum value was identified as $29.3$\,s for the $g$ filter, $31.7$\,s for the $u$ filter and $31.2$\,s for the $r/i$ filter. Therefore, the optimum $\sigma$ for each filter was fixed for the remainder of the study. We identified bad peaks as those which had a maximum relative flux lower than the flux of data points a distance of $2\sigma$ away from them, and we removed candidate peaks that were at the edge of the window. This avoids the inclusion of any peaks where the cycle count is not unambiguously determined. If multiple peaks still existed in a single window after this procedure, the true peak was chosen as the one which was closest to the predicted arrival time. Uncertainties in the arrival times were calculated using Monte-Carlo methods \citep[]{TheSacredText} combined with the exposure time of the observation night. 

Once arrival times had been identified, we used the quadratic ephemeris to assign cycle numbers to each peak, making sure to round the number of elapsed cycles to an integer value. At this point, the arrival time closest to the centre of the data set was assigned as having an integer cycle number of zero in order to minimise the correlation between the period and the period derivative. The correlation between these two variables was analysed using a Markov Chain Monte Carlo (MCMC) fit, and the corresponding corner plot is displayed in Fig.~\ref{fig:CornerPlot}. We then used a Levenberg--Marquardt algorithm \citep[]{levenberg1944method, marquardt1963algorithm}, implemented via the curve\_fit function \citep[]{vugrin2007confidence} from the Scipy library\footnote{https://docs.scipy.org/doc/scipy/index.html}, to fit a new quadratic ephemeris to the arrival times. The residuals folded onto the orbital period of \cite{2023NatAs...7..931P} are displayed in Fig.~\ref{fig:FourierSeries} for the $g$ filter and show clear orbital modulation. Originally, we expected to see a pattern in the residuals representative of the change in the light travel time as the two binary components orbit one another. However, the greatest change in observed minus calculated (O--C) is $\approx36\times$ greater than the expected change due to the light travel time. We discuss potential reasons for this in Section~\ref{sec:orbital modulation}. Nevertheless, we modelled these residuals using a 10-term Fourier series (displayed as a red line in Fig.~\ref{fig:FourierSeries}), and we included this modulation in all future ephemerides. It should be noted that, to model these residuals, we were required to increase our window size to $180$\,s ($\approx0.57\times$ the WD spin period) to capture the entire residuals pattern.  

After fitting a new quadratic ephemeris to the arrival times which incorporated the residuals in Fig.~\ref{fig:FourierSeries}, we updated the trial ephemeris used to predict the locations of the peaks. This meant that we iteratively predicted the locations of the pulse peaks using the trial ephemeris, fitted a quadratic ephemeris to the identified pulse peaks and updated the trial ephemeris with the fitted coefficients until consecutive ephemerides showed no change outside of their uncertainties, ensuring self-consistency. We also had to account for the fact that, like AR Sco, J1912-4410 exhibits some orbital phase values with large O--C scatter \citep[]{2022MNRAS.516.5052P} (even when the residuals in Fig.~\ref{fig:FourierSeries} have been corrected for). This may be due to the emitting region not being directly in our line of sight for some orbital phases. Therefore, each iteration involved two fits: the first allowed points with a large O--C scatter to be masked from the second fit, and then the second fit was used to update the trial ephemeris. Data points were masked if they deviated from the first fitted ephemeris by more than $30$\,s, chosen by trial and inspection of the data.

Once the ephemeris had converged, our final ephemeris for the $g$ filter is:

\begin{equation}
\begin{aligned}
    T_{\text{spin max}}(\text{MJD})=\,\,&60432.170317(8)+0.00369680055(5)E \\&+7.63(15)\times10^{-15}E^2,
\end{aligned}
\label{eq:Ephemeris}
\end{equation}

\noindent where $E$ is the integer cycle number and $T_{\text{spin max}}$ is the arrival time of pulses. Uncertainties are the result of bootstrapping. The final ephemerides of all other filters were in excellent agreement with equation~(\ref{eq:Ephemeris}), with the exception of the $i$ filter, which we discuss in greater detail in Section~\ref{sec: I-filter}. We do not combine the results of different filters into one ephemeris because of the higher errors associated with the $u$ and $r$ filters, discussed in Section~\ref{Section: Observations}. Furthermore, like AR Sco, arrival times of pulses may vary depending on the filter being used, something we also discuss in Section~\ref{sec: I-filter}. Since the quadratic coefficient is equal to $\frac{1}{2}\bar{P}_{\text{spin}}\dot{P}_{\text{spin}}$, where $\bar{P}_{\text{spin}}$ is the average spin period and $\dot{P}_{\text{spin}}$ is the spin period derivative, this leads to $\dot{P}_{\text{spin}}=4.13(8)\times10^{-12}$\,s\,s$^{-1}$. Alternatively, the spin frequency derivative, $\dot{\nu}_{\text{spin}}$, can be expressed as:

\begin{equation}
\begin{aligned}
    \dot{\nu}_{\text{spin}}=\frac{-\dot{P}_{\text{spin}}}{P_{\text{spin}}^2},
\end{aligned}
\label{eq:nudot}
\end{equation}

\noindent yielding $\dot{\nu}_{\text{spin}}=-4.05(8)\times10^{-17}$\,Hz\,s$^{-1}$. The quadratic component of the best-fit ephemeris for the $g$ filter is displayed in Fig.~\ref{fig:OminusCPlot}.

Finally, for each filter with a valid quadratic ephemeris (i.e. $u$, $g$ and $r$), we repeated the entire iterative fitting procedure using a linear ephemeris \citep[our starting linear ephemeris was that from][]{2023NatAs...7..931P} and obtained a best linear ephemeris once the fit had converged. This allowed us to conduct a statistical F-test \citep[]{Schumacker2013} to determine whether the linear ephemeris (the null hypothesis) could be rejected in favour of the quadratic ephemeris (the alternative hypothesis). Our F-tests in all cases found sufficient evidence to reject the null hypothesis, indicating that our quadratic ephemeris was the better fit to the data.

\section{Discussion}
\label{Section: Discussion}

\subsection{Is J1912-4410 Powered by its Spin-Down?}
\label{sec:spin-down}

With our estimate for the spin-down of the WD in J1912-4410, $\dot{P}_{\text{spin}}$, we are now able to calculate the corresponding spin-down luminosity to determine whether this can power the WD pulsar's non-thermal emission, as is the case for AR Sco. The beat period derivative for AR Sco is $\dot{P}_{\text{beat, AR Sco}}=7.18(45)\times10^{-13}$\,s\,s$^{-1}$, and the corresponding spin-down timescale is $t_{\text{AR Sco}}=5.2\pm0.3$\,Myr \citep[]{2018AJ....156..150S}. Once again, we state $\dot{P}_{\text{beat}}$ rather than $\dot{P}_{\text{spin}}$ since the dominant frequency for AR Sco at the time of discovery was the beat frequency \citep[]{2016Natur.537..374M}. Meanwhile, J1912-4410 has a shorter spin-down timescale compared to AR Sco of $t=2.45\pm0.05$\,Myr, which is largely due to the fact that J1912-4410 has a spin-down $\approx6\times$ greater than AR Sco. Therefore, we would expect J1912-4410 to have a higher spin-down luminosity compared to AR Sco as well. 

To calculate the exact spin-down luminosity, we require the standard formula for the rate of loss of rotational energy:

\begin{equation}
\begin{aligned}
    \dot{E}_{\text{rot}}=-4\pi^2I\nu_{\text{spin}}\dot{\nu}_{\text{spin}},
\end{aligned}
\label{eq:spinDownLuminosity}
\end{equation}

\noindent where $I$ is the moment of inertia of the WD, $\nu_{\text{spin}}$ is its spin frequency and $\dot{\nu}_{\text{spin}}$ is the spin frequency derivative. The moment of inertia can be determined from $I=k^2MR^2$ where $M$ and $R$ are the mass and radius of the WD, respectively, and $k=0.452+0.0853\,\text{log}_{10}(1-M/1.44\,\text{M}_{\odot})$ \citep{1985A&A...148..207R, 2021NatAs...5..648S}. By using estimates for $M$ and $R$ from \cite{2024MNRAS.527.3826P}, as well our values for $\nu_{\text{spin}}=1/P_{\text{spin}}$ and $\dot{\nu}_{\text{spin}}$, we obtain an estimate for the spin-down luminosity of $\approx9^{+6}_{-5}\times10^{25}$\,W, with uncertainties propagated using Monte Carlo methods. Our estimate is approximately equal to the order of magnitude estimate for the bolometric luminosity of J1912-4410 from \cite{2023NatAs...7..931P} of $\approx10^{26}$\,W. However, the true spin-down luminosity could be anywhere from $1\times10^{25}$\,W to $2.2\times10^{26}$\,W, given by the $95\,\%$ confidence interval. Here, the main contributors to this large error are the mass and radius estimates for the WD in the moment of inertia estimate, which have percentage errors of $8$\,\% and $31$\,\% respectively. We calculated a $59\,\%$ probability that the spin-down luminosity is less than $10^{26}$\,W. Therefore, our result cannot rule out a spin-down-powered scenario. However, this is vastly different from AR Sco, which has a spin-down luminosity that is an order of magnitude greater than its bolometric luminosity \citep[]{2016Natur.537..374M}. By contrast, our calculations found an $\approx0$\,\% chance that the spin-down luminosity is greater than $10^{27}$\,W. As a result, this calls into question whether J1912-4410 is powered solely by its spin-down. Such a possibility was first hinted at by \cite{2023NatAs...7..931P}, where potential evidence of an accretion-induced flare was detected, implying the possibility of ongoing mass transfer. Such mass transfer could contribute to the total bolometric luminosity. More accurate estimates for the bolometric and spin-down luminosities of J1912-4410 will be needed to provide a more definitive answer to these matters, since our findings cannot rule out a purely spin-down-powered scenario. However, it appears more likely that J1912-4410 is not solely powered by its spin-down, and therefore alternative sources of energy, such as accretion, may contribute to power its non-thermal emission.

\subsection{Compatibility with the Synchronar Model}
\label{sec:Synchronar}
In Section~\ref{sec:Introduction}, we discussed the use of the \textit{synchronar model} for describing the emission mechanism of AR Sco. Proposed by \cite{2017ApJ...835..150K}, the synchronar model describes how particles from the M-dwarf atmosphere are accelerated by the magnetosphere of the WD, and this produces the non-thermal emission we observe. This means that, for the synchronar model to be valid, the WD must have a sufficiently strong magnetic field to provide the necessary torque on the companion star. Now that we have an estimate for the spin-down of the WD in J1912-4410, we are able to determine whether the synchronar model can explain the system's emission. More specifically, we can determine whether the upper limit on the WD's magnetic field of $\approx50$\,MG from \cite{2024MNRAS.527.3826P} is sufficient to provide the necessary torque. 

To calculate the minimum magnetic field at the WD surface for the synchronar model to be valid, we use equations~(3), (4) and (5) from \cite{2017ApJ...835..150K}. We calculate the moment of inertia from the same method as in Section~\ref{sec:spin-down}, using the mass and radius estimates from \cite{2024MNRAS.527.3826P}. We also obtain the radius of the M-dwarf from \cite{2023NatAs...7..931P}. The orbital separation is calculated using Kepler's 3rd law, where we utilise the orbital period and companion mass from \cite{2023NatAs...7..931P}. Finally, we add in our estimate for the spin frequency derivative, $\dot{\nu}_{\text{spin}}$. From this, we obtain a minimum WD magnetic field strength of $110\pm50$\,MG, greater than the upper limit of $\approx50$\,MG found in \cite{2024MNRAS.527.3826P}. As a result, our findings suggest that the emission of J1912-4410 is unlikely to be explained by the synchronar model.

Both the results from this section and Section~\ref{sec:spin-down} appear to suggest that the emission mechanism and energy source of J1912-4410's non-thermal emission could be different to those of AR Sco. Such a result would mean that WD pulsars are a heterogeneous class of WD binary systems. However, similar investigations into other WD pulsars and LPTs will be needed to show whether such distinctions continue across other class members. 

\subsection{Residuals of Pulse Arrival Times}
\label{sec:orbital modulation}
In this section, we conduct a brief analysis of the residuals presented in Fig.~\ref{fig:FourierSeries}, which were obtained when our fit included only the quadratic ephemeris and no Fourier terms. It could be expected that these residuals would be representative of the change in the light travel time as the two binary components orbit each other. However, there are some aspects of Fig.~\ref{fig:FourierSeries} that show this is not the case. Firstly, the change in O--C between orbital phases $0.5$ and $1.0$ is $\approx100$\,s. For light travel time, this corresponds to a distance of $\approx43$\,R$_{\odot}$, which is $\approx36\times$ the orbital separation (using the same calculation for the orbital separation as in Section~\ref{sec:Synchronar}). Also, radial velocity curves from \cite{2023NatAs...7..931P} would suggest that the orbit is approximately circular, yet the residuals in Fig.~\ref{fig:FourierSeries} are highly asymmetric across the first and second halves of the orbit. As a result, the residuals are unlikely to be due to the light travel time alone. Any effect of the light travel time on the shape of the residuals in Fig.~\ref{fig:FourierSeries} would be hidden due to its smaller O--C variability compared to what is shown in the figure. Similarly, AR Sco also shows unexpectedly large residuals, albeit to a lesser extent. Fig.~4 of \cite{2022MNRAS.516.5052P} shows a maximum change in the O--C of $\approx17$\,s which, when converted to a light travel distance, corresponds to $\approx5.5\times$ the orbital separation \citep[to calculate the orbital separation, we use Kepler's 3rd law with the orbital period and median masses from][]{2016Natur.537..374M}. Therefore, it is possible that the mechanism dictating the shape of the residuals of J1912-4410 is also present in AR Sco.

Turning our attention back to J1912-4410, Fig.~\ref{fig:FourierSeries} uses the same zero point for the orbital phase as the radial velocity curves in Fig.~1a of \cite{2023NatAs...7..931P}, so we can relate the different features of the residuals to different points in the binary orbit. From this figure, the orbital phase $0.5$ corresponds to superior conjunction of the M dwarf and phase $0.0/1.0$ corresponds to its inferior conjunction. This means that the greatest values of O--C occur at inferior conjunction and the lowest values of O--C occur at superior conjunction. 

Inspection of Fig.~\ref{fig:FourierSeries} also shows that the orbital phase pattern is divided into smaller sections, separated by discontinuities. We could go further and say that there are six equally-spaced sections in the orbital phase pattern, occurring at intervals $0.0-0.16$, $0.16-0.34$, $0.34-0.51$, $0.51-0.66$, $0.66-0.84$ and $0.84-1.0$. The divisions of these sections can be seen in the fitted Fourier series as the turning points, as well as the kinks in the rise from minimum to maximum O--C between orbital phases $0.5$ and $1.0$. Furthermore, these six sections per orbital period alternate between slow and fast areas, which are identified by sparsely and densely populated areas of the residuals respectively. For example, orbital phases of $0.0-0.16$, $0.33-0.5$ and $0.66-0.84$ are more sparsely populated compared to the other three sections of the residuals. For phases $0.0-0.16$ and $0.33-0.5$, these involve sharp transitions in O--C value. The residuals pattern itself is independent of the Fourier series we used to model the residuals, but the division of the orbital phase pattern into equally-spaced regions may be exaggerated by the Fourier fit. Nevertheless, it is possible this could point to some periodicity associated with the M dwarf, since we expect some level of interaction between the WD and the M dwarf which triggers the periodic emission \citep[]{2016Natur.537..374M}. However, since the exact area of emission for WD pulsars is uncertain \citep[]{2022MNRAS.516.5052P, 2025ApJ...986...78B}, it is difficult to make any sort of geometric argument to explain these residuals. We can say that, assuming a physical origin of the residuals, these sharp transitions in O--C value cannot be longer than the spin period of the WD, as otherwise we would see data points in these gaps.

One peculiar aspect of the lightcurve of J1912-4410 is that the shapes of the pulsations are not consistent. For example, the pulses can alternate between being single-humped and double-humped. Therefore, we also investigated whether the residuals in Fig.~\ref{fig:FourierSeries} were the result of the data consisting of two periodic signals with a small period difference. To do this, we created artificial data that consisted of the superposition of two Gaussians, repeating at their own respective periods, on top of a flat baseline. Firstly, we set the amplitudes and FWHMs of the Gaussians, as well as the signal baseline, to be typical values of the $g$ filter data: an amplitude of $0.04$, a baseline of $0.01$ (in units of relative flux) and a FWHM of $70$\,s. Each Gaussian started at a position of $0.5\times$ its respective period from time zero. The number of simulated orbits was set to the total number of orbits covered in the $g$ filter data, which was $\approx13$. Random flux noise was also applied to the data, with the noise value for each data point drawn from a Gaussian distribution. The distribution had a mean of $0$ and standard deviation set to the typical size of relative flux error in the $g$ filter data, $0.002$. Included in the artificial data set was also a corresponding flux error value for each data point, which we set as a constant $0.002$. We initially set the periods of each Gaussian to be the spin and beat periods of J1912-4410. To calculate the beat period, we used the spin period from equation~(\ref{eq:Ephemeris}) and the orbital period from \cite{2023NatAs...7..931P}. Every artificial data set was then run through the method outlined in Section~\ref{Section: Data Analysis} to produce the corresponding residuals plot. For the error in arrival times, we included a typical $g$ filter exposure time of $3.8$\,s. We should also note that, in this section, the window size where we searched for the pulse peak had to be increased to $200$\,s to capture the full extent of the residuals. The resulting residuals plot is displayed in Fig.~\ref{fig:ResidualsExperiment}. Despite not reproducing the exact residuals pattern shown in Fig.~\ref{fig:FourierSeries}, these residuals do show strong resemblances. Firstly, the residuals are periodic on the orbital period. Secondly, Fig.~\ref{fig:ResidualsExperiment} shows that it is possible to have variations in O--C that exceed the orbital separation (in terms of light travel distance) from a signal consisting of two repeating Gaussians. Thirdly, these residuals show sharp discontinuities, like in Fig.~\ref{fig:FourierSeries}. Therefore, the results of our investigation suggest that the residuals of J1912-4410 may be the result of two overlapping periodic signals originating from the spin and beat periods of J1912-4410. However, a more complete method of generating artificial data will be needed to verify whether this is the case.

\subsection{$i$-Filter Photometry}
\label{sec: I-filter}

From our analysis of the optical data in Section~\ref{Section: Data Analysis}, we found that the $g$ filter data gave a best-fit ephemeris displayed in equation~(\ref{eq:Ephemeris}), and that the $u$ and $r$ filter data agreed with this ephemeris. On the other hand, for the $i$ filter data, the ephemeris fitting procedure failed significantly more often than for any other filter, and when the fit did converge, the best-fit ephemerides were not consistent with each other, nor with the other filters. One may suggest that the lack of data points in the $i$ filter is the reason for this issue. However, the $r$ filter had even fewer data points (15 nights of data for the $i$ filter compared to 13 nights of data for the $r$ filter, see Table~\ref{tab:Observations}) and was still able to converge to consistent values. Therefore, another possibility is that the arrival times of pulses for J1912-4410 vary depending on the filter being used for observations. Such a phenomenon could arise for a number of reasons: the point of origin of the light contained within each bandpass could be different due to the system's temperature distribution, or there may be material surrounding the system which causes a delay in the light travel time that is wavelength dependent. Subsequently, pulses in the $i$ band could arrive at different times to the other three bandpasses. This is also something that was discovered for AR Sco by \cite{2020MNRAS.496.4849G}, where the arrival times of pulses could differ by as much as $6$\,s between data in the $u$ and $r$ photometric bands. If the arrival times for the $i$ filter data were different compared to the $ugr$ data, poor initial guesses for our ephemeris fitting, combined with the lower number of data points, may have led to the increased number of failed converges and inconsistent estimates for the quadratic ephemeris. More $i$ filter data over a longer timespan will be necessary to determine whether this is the case, however this is something that should be considered when collecting optical data on J1912-4410 for future studies.

\section{Summary \& Conclusions}

We have determined the first estimate for the spin-down of the WD in J1912-4410 using high-speed optical photometric data in multiple bandpasses. We used ULTRACAM observations in the $u$, $g$, $r$ and $i$ bandpasses spanning the last four years, with a total of 28 nights of observation. To determine the best fit ephemeris, we identified pulse arrival times by convolving our data set with a Gaussian function and inspecting where this cross-correlation function was maximised. We used a trial ephemeris to assign integer cycle numbers to each identified pulse location. We then fit a quadratic ephemeris (including a Fourier component to account for the residuals in Fig.~\ref{fig:FourierSeries}) to these cycle numbers, used the fitted ephemeris to update the integer cycle numbers, and iteratively did this until the trial and fitted ephemerides converged.

After completing this procedure, we obtained a best fit ephemeris which gave a spin period derivative of $\dot{P}_{\text{spin}}=4.13(8)\times10^{-12}$\,s\,s$^{-1}$ using $g$ filter data. With the support of a statistical F-test, this confirmed that the WD in J1912-4410 is undergoing spin-down. The result also corresponds to a spin frequency derivative of $\dot{\nu}_{\text{spin}}=-4.05(8)\times10^{-17}$\,Hz\,s$^{-1}$. Our best fit ephemeris is in excellent agreement with our findings from the $u$ and $r$ filter data, yet it disagreed with our findings from the $i$ filter data. This may be due to a lower number of observing nights for the $i$ filter compared to the $u$ and $g$ filters, yet another possibility is that the pulse arrival times of J1912-4410 vary depending on the particular bandpass used for observation. More $i$ filter data over a longer timespan will be needed to determine the true cause of the discrepancy.

From our spin period derivative, we calculated a spin-down luminosity of $9^{+6}_{-5}\times10^{25}$\,W, a value that is approximately equal to the order of magnitude estimate for the bolometric luminosity of J1912-4410 from \cite{2023NatAs...7..931P}. The errors on our estimate mean that we cannot rule out a purely spin-down-powered scenario, like AR Sco. However, our findings indicate that it is likely J1912-4410 is not solely powered by its spin-down, and therefore the WD pulsar may have other contributions, such as accretion, to power its non-thermal emission. Additionally, our spin period derivative allowed us to estimate the minimum magnetic field that the WD requires for the emission of J1912-4410 to be consistent with the \textit{synchronar model}: a model suggested by \cite{2017ApJ...835..150K} to explain the emission of AR Sco. We found a minimum magnetic field strength of $110\pm50$\,MG, which is greater than the experimentally-determined upper limit for the WD's magnetic field of $\approx50$\,MG from \cite{2024MNRAS.527.3826P}. This suggests that the non-thermal emission of J1912-4410 is unlikely to be explainable by the synchronar model, and hence is generated via a different mechanism.

Finally, we found an unusual residuals pattern that is modulated on the orbital period, Fig.~\ref{fig:FourierSeries}, which we modelled as a Fourier series as part of our fit to arrive at our estimate for the spin-down. AR Sco was also found to exhibit unusual residuals \citep{2022MNRAS.516.5052P}, yet the residuals were smaller compared to those from this study. We suspect these residuals may be the result of the data consisting of two periodic signals with a small period difference. However, further work will be necessary to determine whether this is the case and, if so, what physical mechanisms lead to these two signals.

Our findings imply that WD pulsars are a heterogeneous class of WD binary systems. Similar investigations into other WD pulsars and LPTs will be necessary to determine whether this pattern of diversity continues across other members of the class.

\section*{Acknowledgements}

We thank the referee for very helpful scientific insight that improved this manuscript. CJH and IP would like to thank the Royal Astronomical Society for funding this work through an Undergraduate Bursary. IP acknowledges support from the Royal Society through a University Research Fellowship (URF\textbackslash R1\textbackslash 231496). ULTRACAM operations are funded by the Science and Technology Facilities Council (grant ST/Z000033/1). NCS acknowledge support from the Science and Technology Facilities Council (STFC) grant ST/X001121/1. VSD is supported by a Leverhulme Research Fellowship (RF-2025-297/9) and the Spanish Ministry of Science, Innovation and Universities (PID2023-151588NB-I00).

%%%%%%%%%%%%%%%%%%%%%%%%%%%%%%%%%%%%%%%%%%%%%%%%%%
\section*{Data Availability}

All data analysed in this work can be made available upon reasonable request to the authors.



\bibliographystyle{mnras}
\bibliography{export-bibtex} % if your bibtex file is called example.bib

%\begin{thebibliography}{99}
%\end{thebibliography}

%%%%%%%%%%%%%%%%%%%%%%%%%%%%%%%%%%%%%%%%%%%%%%%%%%

%%%%%%%%%%%%%%%%% APPENDICES %%%%%%%%%%%%%%%%%%%%%

\appendix

\section{Change in Orbital Modulation of J1912-4410}
\label{appsec: orbital modulation}

\begin{figure*}
	\includegraphics[width=\textwidth]{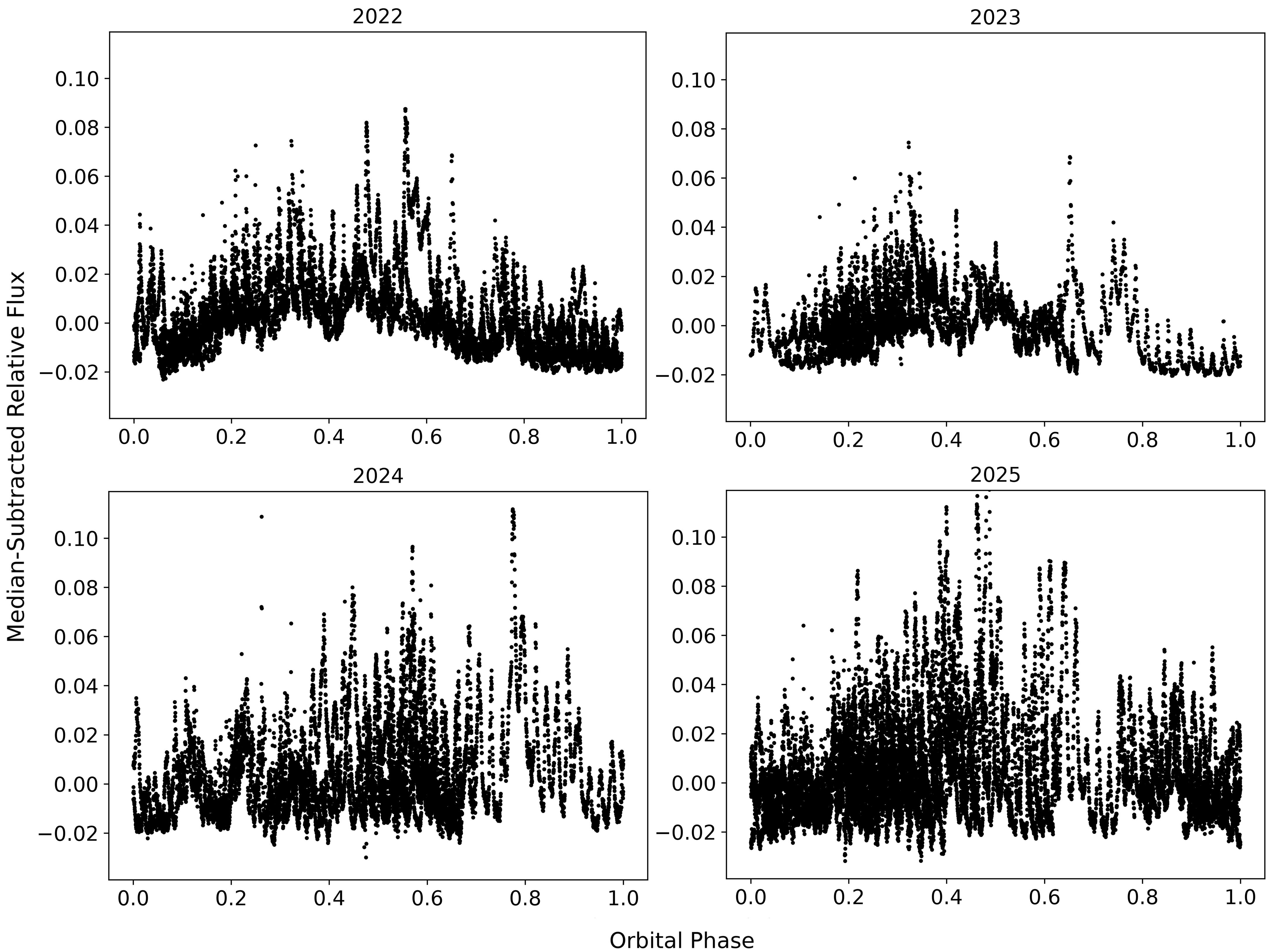}
    \caption{Orbital modulation of median-subtracted relative flux for each year of observation of J1912-4410. Data presented is for the $g$ filter, yet the same orbital modulation was also seen in the $u$ filter. Here, we do not include a graph for the year 2026, because we did not have enough data to cover the entire orbital period.}
    \label{fig:OrbModperYear}
\end{figure*}

In Section~\ref{Section: Data Analysis}, we intended to remove any variations in the lightcurve of J1912-4410 that were modulated on the orbital period. This meant that, for each night of observation, we median-subtracted the data, and then the entire data were orbital phase-folded \citep[using the orbital period of][]{2023NatAs...7..931P} and binned. However, unexpectedly, we found that the shape of the underlying orbital modulation changed across the four years that observations were taken (see Fig.~\ref{fig:OrbModperYear}). It should be noted that our best quadratic ephemeris, equation~(\ref{eq:Ephemeris}), does not change regardless of if we remove the orbital modulation for the entire data at once or if we remove the orbital modulation separately for each year. For this reason, we largely follow the method described in Section~\ref{Section: Data Analysis} for our study. It is unclear what precisely is causing this change to the orbital modulation, but it is something that should be kept in consideration for future photometric studies of J1912-4410.

\section{Corner Plot from MCMC Fit}

\begin{figure*}
	\includegraphics[width=\textwidth]{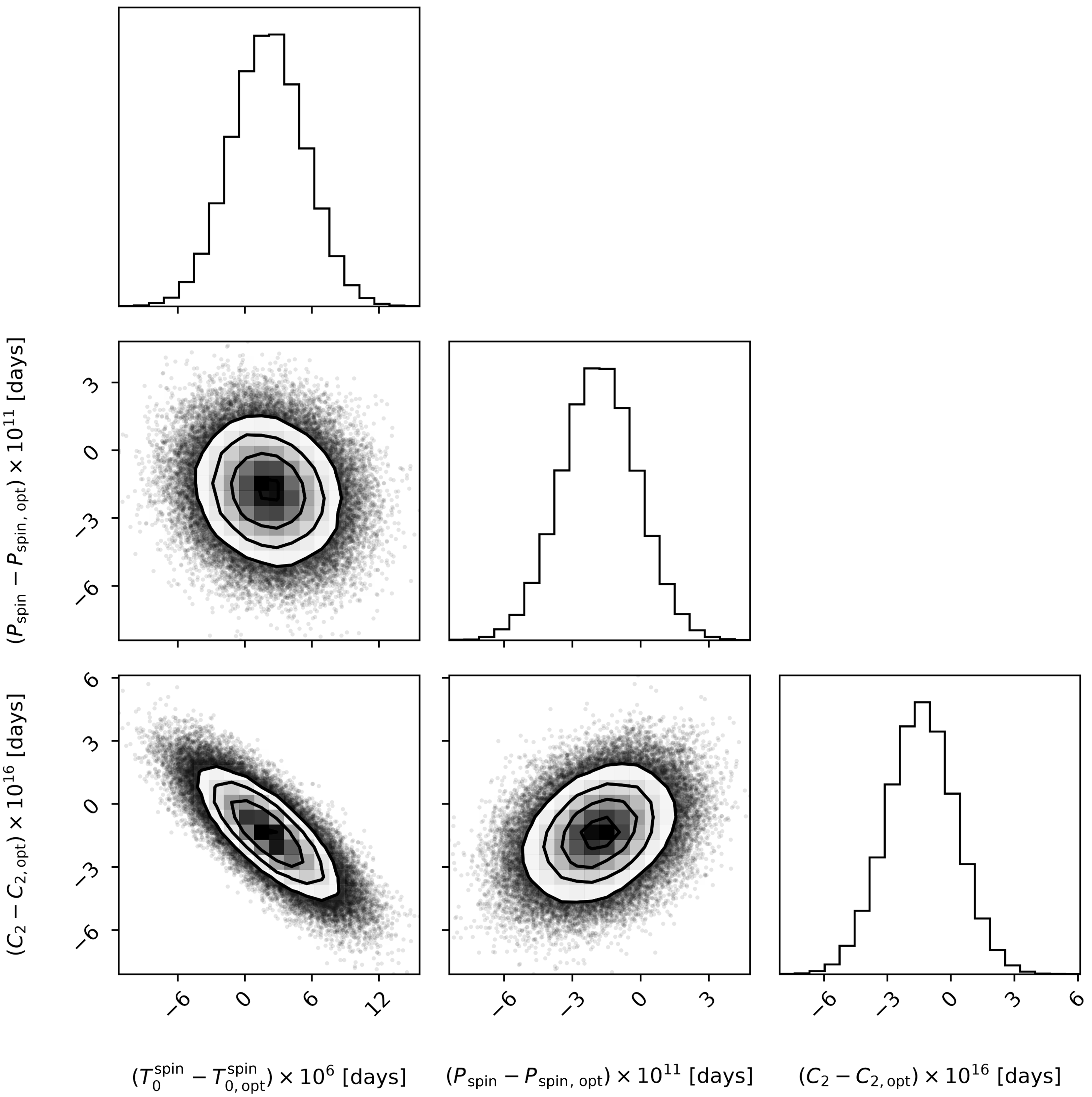}
    \caption{Corner plot from the MCMC fit of the $g$ filter, displaying the three coefficients from the quadratic ephemeris and their correlation with each other. Here, $T^{\text{spin}}_{\text{0}}$ is the time with an integer cycle number of zero, $P_{\text{spin}}$ is the spin period of J1912-4410 and $C_{2}$ is the quadratic coefficient of the ephemeris. For each variable, we have subtracted the best-fit values from equation~(\ref{eq:Ephemeris}), which we denote using the \textit{opt} subscript. All median MCMC values are in excellent agreement with the best-fit ephemeris.}
    \label{fig:CornerPlot}
\end{figure*}

\section{Residuals Plot using Artificial Data}

\begin{figure}
	\includegraphics[width=\columnwidth]{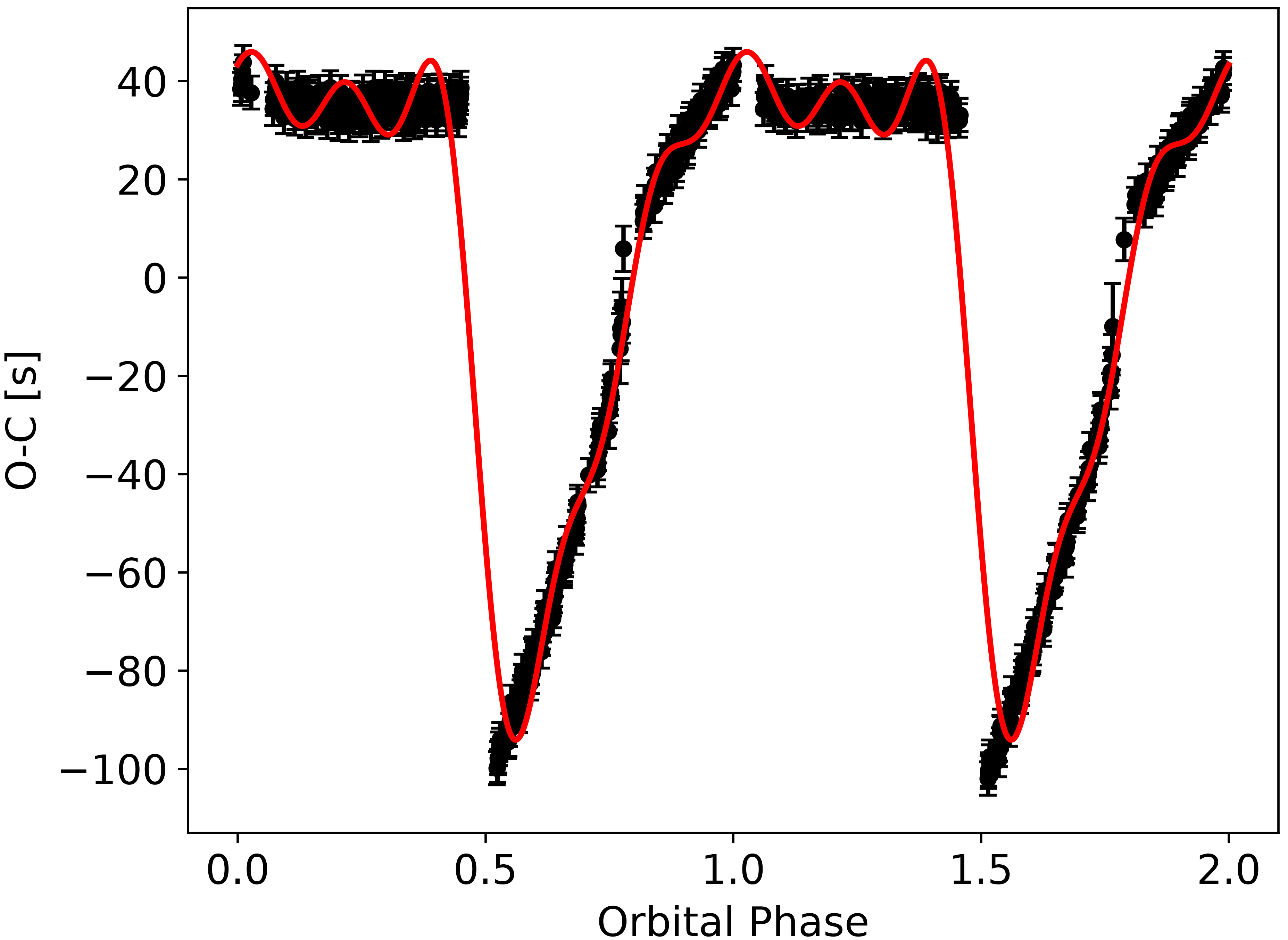}
    \caption{Observed minus calculated (O--C) residuals of identified pulse arrival times for the artificial data described in Section~\ref{sec:orbital modulation}, phase-folded onto the orbital period. The red line displays a 10-term Fourier series that is used to model this behaviour.}
    \label{fig:ResidualsExperiment}
\end{figure}

\bsp
\label{lastpage}
\end{document}